\documentclass[twocolumn,aps,prd]{revtex4}
 \usepackage{graphicx,amssymb}
 \usepackage[utf8]{inputenc} 
 
\usepackage[urlcolor=blue,colorlinks,breaklinks=true]{hyperref}

\PassOptionsToPackage{obeyspaces,spaces,hyphens}{url}
 
 \usepackage{amsmath}

\begin{document}
 	\def\half{{1\over2}}
 	\def\shalf{\textstyle{{1\over2}}}
 	
 	\newcommand\lsim{\mathrel{\rlap{\lower4pt\hbox{\hskip1pt$\sim$}}
 			\raise1pt\hbox{$<$}}}
 	\newcommand\gsim{\mathrel{\rlap{\lower4pt\hbox{\hskip1pt$\sim$}}
 			\raise1pt\hbox{$>$}}}

\newcommand{\be}{\begin{equation}}
\newcommand{\ee}{\end{equation}}
\newcommand{\bq}{\begin{eqnarray}}
\newcommand{\eq}{\end{eqnarray}}
 	

\title{Can gravitational vacuum condensate stars be a dark energy source?}
 	 	
\author{P.P. Avelino}
\email[Electronic address: ]{pedro.avelino@astro.up.pt}
\affiliation{Departamento de F\'{\i}sica e Astronomia, Faculdade de Ci\^encias, Universidade do Porto, Rua do Campo Alegre 687, PT4169-007 Porto, Portugal}
\affiliation{Instituto de Astrof\'{\i}sica e Ci\^encias do Espa{\c c}o, Universidade do Porto, CAUP, Rua das Estrelas, PT4150-762 Porto, Portugal}
\affiliation{Centro de Astrof\'{\i}sica da Universidade do Porto, Rua das Estrelas, PT4150-762 Porto, Portugal}

\date{\today}
\begin{abstract}
	
Gravitational vacuum condensate stars, also known as gravastars, have been proposed as an alternative to black holes. Their interior contains a perfect fluid with an equation of state akin to that of a cosmological constant. For this reason, they have recently been considered as a possible astrophysical source of dark energy. In this work we argue that gravitational vacuum condensate stars cannot be the source of dark energy and highlight that a direct coupling of their mass to the dynamics of the Universe would lead to an additional velocity dependent acceleration, damping their motion with respect to the cosmological frame. We briefly discuss the potential impact of this additional acceleration in the context of a recent proposal that the observed mass growth of compact objects at the core of elliptical galaxies might result from such a cosmological coupling.
		
\end{abstract}

\maketitle
 	
\section{Introduction}
\label{sec:intr}

Gravitational vacuum condensate stars \cite{Dymnikova:1992ux,Mazur:2001fv,Mazur:2004fk,Lobo:2005uf} (see \cite{Mottola:2023jxl} for a recent review), also known as gravastars, are compact objects whose interiors are essentially described by a perfect fluid with a dark energy equation of state. It has been argued that the pressure inside gravastars may be responsible for the acceleration of the expansion of the Universe, thus providing a possible astrophysical source of dark energy \cite{Croker:2019mup,2020ApJ...889..115C,2020ApJ...900...57C} (see, however, \cite{Parnovsky:2023wkc} for a different perspective). It has also been suggested that this should be associated to a corresponding growth of the mass of gravastars, which could be constrained observationally. In fact, recent observations of elliptical galaxies with a redshift $z \lsim 2.5$ \cite{2023ApJ...943..133F,Farrah:2023opk} have been interpreted as evidence for a cosmological coupling of the mass of these objects, with a zero cosmological coupling claimed to be excluded at $99.98 \%$ confidence (see, however, \cite{2020ApJ...889...32S,Rodriguez:2023gaa} for an alternative view).

In this work we analyse whether the pressure inside stable compact objects may potentially affect the expansion of the Universe and shed light on the nature of the dark energy. We shall essentially assume energy-momentum conservation and Birkhoff's theorem in general relativity \cite{1921ArMAF..15...18J,1923rmp..book.....B}, considering first the simpler case of stable compact objects with negligible self-gravity. We also discuss possible implications of energy-momentum conservation to the dynamics of these objects in a cosmological scenario where the growth of the mass of compact objects, such as gravastars, is directly coupled to the dynamics of the Universe. 

Except when stated otherwise, we shall use units such that $c=1$ ($c$ is the speed of light in vacuum).

\section{The pressure inside stable compact objects and the expansion of the Universe}
\label{sec:sec1}

The volume-averaged pressure inside stable static compact objects with a negligible self-induced gravitational field vanishes. A general derivation of this condition, also known as the von Laue condition \cite{doi:10.1002/andp.19113400808},  can be found in  \cite{Avelino:2018qgt} --- see also \cite{Avelino:2018rsb,Ferreira:2020fma,Avelino:2022eqm} for a detailed account of how this generic result is related to the appropriate form of the on-shell Lagrangian of an ideal gas.

Here, we consider an alternative derivation of the von Laue condition using the standard energy-momentum conservation law, 
\be
\frac{d \bar \rho}{dt}=-3H(\bar \rho+\bar p)=-3H \bar \rho(1+w) \label{firstlaw}\,,
\ee
in an expanding homogeneous and isotropic Friedmann-Lemaître-Robertson-Walker universe permeated with a perfect fluid ($\bar \rho$ and $\bar p$ are, respectively, the cosmological energy density and pressure, $w=\bar p/\bar \rho$ is the cosmological equation of state parameter, $H={\dot a}/a$ is the Hubble parameter, $a$ is the cosmological scale factor and a dot represents a derivative with respect to the physical time $t$). The cosmological energy density of a homogeneous and isotropic population of stable compact objects comoving with the expansion of the universe evolves as $\bar \rho \propto a^{-3}$, and Eq.~(\ref{firstlaw}) implies that $\bar \rho \propto a^{-3}$ if and only if $\bar p=0$. Assuming that the cosmological pressure $\bar p$ in Eq.~(\ref{firstlaw}) is the volume average of all pressures everywhere \cite{Croker:2019mup}, we find that the volume-averaged pressure inside stable static compact objects must vanish (otherwise their contribution to the cosmological pressure $\bar p$ would not vanish and their cosmological energy density would not evolve as $\bar \rho \propto a^{-3}$). 

If the self-induced gravitational field cannot be neglected, then the von Laue condition no longer applies. As an example, let us consider a stable non-relativistic spherically compact object (e.g. a star) of radius $R$ and mass $m \equiv m[R] \ll R/G$ in hydrostatic equilibrium. In this limit the pressure field $p[r]$ satisfies 
\be
p'[r]\equiv\frac{dp[r]}{dr}=-\frac{G m[r] \rho[r]}{r^2}\,, \label{hydrostatic}
\ee
where
\be
m[r]=4 \pi \int_0^{r} \rho[u] u^2 du
\ee
is the mass inside a radius $r$ and $\rho[r]$ represents the matter density as a function of the distance to the center of the compact object. The volume-averaged density and pressure are given, respectively, by
\bq
\langle \rho \rangle &=& \frac{3m}{4\pi R^3}\,,\\ 
\langle p \rangle &=& \frac{\int_0^{R} p[r] r^2 dr}{\int_0^{R} r^2 dr}=-\frac{\int_0^{R} p'[r] r^3 dr}{R^3} = -\frac{U_{\rm g}}{4\pi R^3} \nonumber \,,
\eq
where
\be
U_{\rm g} = -4 \pi G \int_0^{R} m[r] \rho[r] r dr \sim -\frac{G m^2}{R} \,,
\ee
is the total gravitational potential energy.
Hence, the average pressure inside the compact object is given by $\langle p\rangle = w_{\rm co} \langle \rho \rangle$,  with
\be
w_{\rm co} =- \frac{U_{\rm g}}{3 m} \sim \frac{G m}{R} \neq 0\,.
\ee
As predicted by the von Laue condition, the pressure vanishes in the $G m/R \to 0$ limit. On the other hand, for a fixed mass $m \equiv m[R]$ the average pressure is expected to be roughly proportional to $R^{-1}$. One may now raise the question of whether the expansion of the Universe can be sensitive to the distribution of radii $R$ of stable compact objects with $R \ll H^{-1}$, assuming that their contribution to the cosmological energy density is fixed. In other words, could the values of $w$ and $w_{\rm co}$ be related?

Although the von Laue condition cannot be applied to objects with strong self-induced gravitational fields, according to Birkhoff's theorem \cite{1921ArMAF..15...18J,1923rmp..book.....B} in general relativity exterior spherically symmetric vacuum solutions are described by the Schwarzschild metric independently of the strength of the self-induced gravitational field. Hence, the interior pressure does not contribute to the gravitational field outside spherically symmetric stable compact objects. As a result, it also cannot not affect the expansion of the Universe. This is true even if the interior pressure is extraordinarily high as happens, for example, in the case of neutron stars or (hypothetical) gravastars. Therefore, gravitational vacuum condensate stars cannot be a source of dark energy.  

It has been shown in \cite{Croker:2019mup} that, under certain assumptions, the cosmological pressure $\bar p$ in Eq.~(\ref{firstlaw}) is the volume average of all pressures everywhere, including the interiors of compact objects with strong gravitational fields. However, this result (Eqs. (44) and (54) in \cite{Croker:2019mup}) relies on perturbation theory with respect to a homogeneous and isotropic Friedmann-Lemaitre-Robertson-Walker background and neglects the local gravitational field which balances the pressure field inside stable compact objects with non-negligible volume-averaged pressure. The local gravitational field is crucial in this context, since it effectively cancels the contribution of the local pressure inside stable compact objects to the cosmological pressure.

\section{Mass growth and energy-momentum conservation}
\label{sec:sec2}

The cosmological density and pressure associated to a homogeneous and isotropic distribution of stable compact objects of constant proper mass $m_0$ and speed $v$ is given by $\bar \rho\propto m_0 \gamma a^{-3}$ and $\bar p=\bar \rho v^2/3$. Substituting into Eq. (\ref{firstlaw}) assuming a constant $m_0$ it is straightforward to show that
\be
\frac{d (\gamma v)}{dt} + H \gamma v = 0 \label{velocity}\,,
\ee
or equivalently, that the value of the linear momentum of the compact objects with respect to the cosmological frame evolves as $m_0 \gamma v \propto a^{-1}$ [here $\gamma=(1-v^2)^{-1/2}$]. 

The simplest way to couple the proper mass $m_0$ of a compact object to the dynamics of the Universe is to consider a model where the proper mass is a function of a (nearly) homogeneous dark energy scalar field $\phi=\phi[a]$ \cite{Farrar:2003uw,2012PhRvD..85l3010A,Avelino:2015fka}. In fact, the scalar field cannot be perfectly homogeneous because otherwise there would be no energy flow and $m_0$ would remain a constant. However, for simplicity, in the following we shall approximate the scalar field as being perfectly homogeneous, thus neglecting the associated fifth forces.

According to general relativity the energy-momentum tensor must be covariantly conserved and, therefore, the increase of the proper mass of the compact object $m_0[a]$ needs to be compensated by a corresponding decrease of its speed $v$ with respect to the local cosmological frame. Hence, in order to account for a direct coupling of the proper mass $m_0$ to the scale factor $a$, one needs to add a momentum conserving source term to the right hand side of Eq. (\ref{velocity}) satisfying
\be
\frac{d \left(m_0 \gamma v\right)}{dt}=0\,, \label{source}
\ee
or equivalently
\be
\left[\frac{d (\gamma v)}{dt}\right]_{\rm source} = - \frac{d \ln m_0}{dt} \gamma v = -k H \gamma v\,, \label{source1}
\ee
with
\be
k = \frac{d (\ln m_0)}{d (\ln a)}\,.
\ee
With this added term, Eq. (\ref{velocity}) now becomes
\be
\frac{d (\gamma v)}{dt} + H \left(1+k\right) \gamma v = 0 \label{velocity1}\,.
\ee
Hence, the mass growth of the compact objects would be associated with a larger damping of their velocity by a factor of $1+k$. For $k=\rm const$, Eq. (\ref{velocity1}) implies that $\gamma v \propto a^{-1-k}$ in a perfectly homogeneous and isotropic Friedmann-Lemaître-Robertson-Walker universe, instead of the standard result $\gamma v \propto a^{-1}$. 

\section{Acceleration of gravastars with respect to their host galaxies}
\label{sec:sec3}

In the previous section we have shown that a direct coupling of the growth of the proper mass $m_0$ of a gravastar to the scale factor $a$ must be associated to a velocity dependent acceleration causing an additional damping of its velocity. According to Eq. (\ref{velocity1}), if the gravastar motion with respect to the local cosmological frame is non-relativistic then the corresponding acceleration with respect to a comoving object whose mass is not coupled to the evolution of the Universe should be equal to
\be
\dot {\vec v} _{\rm coupling}= \frac{d \vec v}{dt} = - k H \vec v  \,.
\ee

The speed of our own galaxy, the Milky Way, with respect to the Cosmic Microwave Background has been estimated to be equal to $631 \pm 20 \, \rm km \, s^{-1}$ \cite{2017NatAs...1E..36H}. Using $H_0 = 70 \, \rm km \, s^{-1} \, Mpc^{-1}$, and assuming the value of the cosmological coupling strength suggested by the observational analysis in \cite{Farrah:2023opk} ($k=3$),  one can estimate that the value of $\dot {\vec v} _{\rm coupling}$ at the present time would be $|\dot {\vec v} _{\rm coupling0}|=4.3 \times 10^{-12} \, \rm m  \, s^{-2}$ for a gravastar at the core of Milky Way. Although this is roughly only $2 \%$ of the acceleration of the Solar System barycenter with respect to the rest frame of the Universe (measured from Gaia astrometry to be equal to $2.32 \pm  0.16 \times 10^{-10} \, \rm m  \, s^{-2}$ \cite{2021A&A...649A...9G}), the cosmological coupling of the proper mass gravastars could potentially  be responsible for a shift in the position of gravastars away from the center of their host galaxies, with the amplitude of the shift being dependent on the relative velocity of each galaxy with respect to the local cosmological frame. The dynamical impact of a cosmological coupling of the proper mass gravastars could be more significant at larger redshifts, since the linear growth of cosmological perturbations is particularly affected by modifications to the cosmological damping term.

\section{Conclusions}\label{sec:conc}

The precise nature of the dark sector of the Universe remains elusive. Therefore, it is interesting to explore different avenues for what could be the source of dark matter and dark energy. In this work we analysed a recent suggestion that gravastars could potentially be the source of dark energy. 

We started by providing an alternative derivation of the von Laue condition, which states that the volume-averaged pressure inside stable compact objects with a negligible self-induced gravitational field vanishes, not contributing to the cosmological pressure. We argued that, although the von Laue condition cannot be applied in the presence of a significant self-induced gravitational field, Birkhoff's theorem in general relativity precludes the pressure inside stable compact objects from playing a cosmological role, thus excluding gravastars as a plausible dark energy source. We have also highlighted that the local gravitational field which balances the pressure field  inside stable compact objects with non-negligible volume-averaged pressure must be considered when performing cosmological averages, since it effectively cancels the contribution of the local pressure inside stable compact objects to the cosmological pressure.

Energy-momentum conservation implies that a direct coupling of the proper mass of stable compact objects to the cosmological scale factor must lead to an additional velocity dependent acceleration, damping their motion with respect to the cosmological frame. We have estimated this acceleration for a gravastar at the core of our own galaxy, assuming the value of the cosmological coupling strength suggested by a recent observational analysis, showing that it is about $2 \%$ of the acceleration of the Solar System barycenter with respect to the rest frame of the Universe. Although this is a relatively small acceleration, it could potentially lead to a shift in the position of gravastars away from the center of their host galaxies. We have also pointed out that the dynamical impact of a cosmological coupling of the proper mass of gravastars could be more prominent at high redshift.
	
\begin{acknowledgments}

We thank Lara Sousa, Rui Azevedo, Vasco Ferreira, and D. Gr{\"u}ber for many enlightening discussions on topics related to this work. We also thank José Afonso for an excellent presentation of his work on a special session of the CosmoClub at Instituto de Astrofísica e Ciências do Espaço and my colleagues in the Cosmology group for the open discussion following that presentation. We acknowledge the support by Fundação para a Ciência e a Tecnologia (FCT) through the research grants UIDB/04434/2020, UIDP/04434/2020. This work was also supported by FCT through the R$\&$D project 2022.03495.PTDC - Uncovering the nature of cosmic strings.

\end{acknowledgments}
 
\bibliography{gravastars}
 	
 \end{document}